\documentclass[a4paper,12pt]{article}
\usepackage[pctex32]{graphics}

\begin{document}
\topmargin 0pt \oddsidemargin 0mm

\renewcommand{\thefootnote}{\fnsymbol{footnote}}
\begin{titlepage}
\begin{flushright}
INJE-TP-05-01\\
hep-th/0501023
\end{flushright}

\vspace{5mm}
\begin{center}
{\Large \bf Cosmic holographic bounds with UV and IR cutoffs}
\vspace{12mm}

{\large  Yun Soo Myung\footnote{e-mail
 address: ysmyung@physics.inje.ac.kr}}
 \\
\vspace{10mm} {\em  Relativity Research Center and School of
Computer Aided Science, Inje University Gimhae 621-749, Korea}
\end{center}

\vspace{5mm} \centerline{{\bf{Abstract}}}
 \vspace{5mm}
We introduce  the cosmic holographic  bounds with two  UV and IR
cutoff scales, to deal with  both the inflationary universe in the
past and dark energy in the future. To describe quantum
fluctuations of inflation on sub-horizon scales, we  use the
Bekenstein-Hawking energy bound. However, it is not justified that
the D-bound is satisfied with the coarse-grained entropy.  The
Hubble bounds are introduced for classical fluctuations of
inflation on super-horizon scales. It turns out that the Hubble
entropy bound is satisfied with the entanglement entropy and the
Hubble temperature bound leads to a condition for the slow-roll
inflation.  In order to describe the dark energy, we introduce the
holographic energy density which is the one saturating the
Bekenstein-Hawking energy bound for a weakly gravitating system.
Here the  UV (IR) cutoff is given by  the Planck scale (future
event horizon), respectively. As a result, we find  the close
connection between quantum and classical fluctuations of
inflation, and dark energy.
\end{titlepage}

\newpage
\renewcommand{\thefootnote}{\arabic{footnote}}
\setcounter{footnote}{0} \setcounter{page}{2}
\section{Introduction}
Supernova (SN Ia) observations\cite{SN} suggest that our universe
is accelerating and the dark energy contributes $\Omega_{\rm
DE}\simeq 0.60-0.70$ to the critical density of the present
universe. Also the cosmic microwave background (CMB) observations
\cite{Wmap} imply that the standard cosmology is given by the
inflation\cite{Inf} and flat FRW universe.  It is generally
accepted that curvature (scalar) perturbations produced during
inflation are considered to be the origin of inhomogeneities
necessary for explaining  CMB anisotropies and large-scale
structures.

The holographic principle is basically a statement about the
number of fundamental degrees of freedom\cite{Beke}. In the case
of scalar perturbations of inflation with the UV cutoff $\Lambda$,
the holographic principle was used to derive the entropy bound for
quantum fluctuations of inflation on sub-horizon scales~\cite{KS}.
It was shown that the indirect entropy bound leads to a constraint
which states that the UV cutoff  is nearly of  the same order of
the Planck scale. Also it was suggested that the D-bound is
satisfied with the coarse-grained and entanglement entropies.

One of promising candidates for the  dark energy is the
cosmological constant.  Cohen {\it et al.}~\cite{CKN} showed that
in quantum field theory, the UV cutoff $\Lambda$ is related to the
IR cutoff $L_{\rm \Lambda}$ because of the limit set by forming a
black hole. In other words, if $\rho_{\rm \Lambda}$ is the quantum
zero-point energy density caused by the UV cutoff, the  energy of
the system with size $L_{\rm \Lambda}$ should not exceed the mass
of the system-size black hole: $L_{\rm \Lambda}^3 \rho_{\rm
\Lambda}\le L_{\rm \Lambda} M_p^2$ with the Planck mass of
$M_p^2=1/G$. The largest $L_{\rm \Lambda}$ is then given by the
one saturating this inequality and the  holographic energy density
takes the form of $\rho_{\rm \Lambda} \sim (M_p^2/L_{\rm
\Lambda})^2$.  Recently the holographic energy density has been
used to explain the dark energy
widely~\cite{HMT,HSU,WM,HOV,LI,FEH,Myung2}. However, we mention
that the future event horizon as the IR cutfof seems to be rather
ad hoc chosen and it thus requires establishing a close connection
between the holographic energy density and dark energy.

On the other hand, the implications of the cosmic holographic
bounds have been investigated in the
literature~\cite{Hooft,FS,Hubb,Bous,Verl,SV}. The cosmic
holographic bounds are based on the holographic principle and the
two Friedmann equations.

In this work we will examine  how the cosmic holographic bounds
with the UV and IR cutoffs could be used for describing
fluctuations of inflation in the past and the  dark energy in the
future.  Explicitly we use the Bekenstein-Hawking energy bound
$E\le E_{\rm BH}$ for a weakly gravitating system with $a\le
\sqrt{2-k}/H$, whereas for a strongly gravitating system with
$a\ge \sqrt{2-k}/H$, we use the Hubble entropy bound  $S \le
S_{\rm H}$ and temperature bound $T\ge T_{\rm H}$. Especially, the
energy bound is very useful for describing quantum fluctuations of
inflation in the past and the dark energy in the future. In these
cases, the UV cutoff is the Planck scale and the IR cutoff scale
is the Hubble horizon for quantum fluctuations of inflation
 and the future event horizon for the dark
energy. For classical fluctuations of inflation on super-horizon
scales, we use the Hubble bounds to estimate the lower limit of
its entropy and the upper limit of its temperature. Here the UV
cutoff scale is also given by the Planck scale and the IR scale is
the Hubble horizon on super-horizon scales.

The organization of this work is as follows. In section 2 we
briefly review the cosmic holographic bounds: the
Bekenstein-Hawking energy bound for a weakly gravitating system
and the Hubble bounds for  a strongly gravitating system. We apply
the cosmic holographic bounds to the fluctuations of inflation:
quantum fluctuations on sub-horizon scales and classical
fluctuations on super-horizon scales in section 3. Section 4 is
devoted to deriving the holographic energy density from the
Bekenstein-Hawking energy bound. We show how the holographic
energy density can explain the dark energy in Section 5. Finally
we summarize our results in section 6.

\section{Cosmic holographic bounds}
We briefly review the cosmic holographic bounds. Let us start an
$(n+1)$-dimensional Friedmann-Robertson-Walker (FRW) metric
\begin{equation}
\label{2eq1} ds^2 =-dt^2 +a(t)^2 \Big[ \frac{dr^2}{1-kr^2} +r^2
d\Omega^2_{n-1} \Big],
\end{equation}
where $a$ is the  scale factor of the universe and
$d\Omega^2_{n-1}$ denotes the line element of an
$(n-1)$-dimensional unit sphere. Here $k=-1,~0,~1$ represent that
the universe  is open, flat, closed, respectively. A cosmological
evolution is determined by the two Friedmann equations
\begin{eqnarray}
\label{2eq2}
 && H^2 =\frac{16\pi G_{n+1}}{n(n-1)}\frac{E}{V}
-\frac{k}{a^2}, \nonumber \\
&& \dot H =-\frac{8\pi G_{n+1}}{n-1}\left (\frac{E}{V} +p\right)
    +\frac{k}{a^2},
\end{eqnarray}
where $H=\dot{a}/a$ represents the Hubble parameter  and the
overdot stands for  derivative with respect to the cosmic time
$t$,  $E$ is the total energy of matter filling the universe, and
$p$ is its pressure. $V$ is the volume of the universe, $V=a^n
\Sigma^n_k$ with $\Sigma^n_k$ being the volume of an
$n$-dimensional space with a curvature constant $k$, and $G_{n+1}$
is the Newton constant in ($n+1$) dimensions.  It denotes
$G=1/M_p^2$ for (3+1)-dimensional spacetime.  Here we assume the
equation of state:
 $p=\omega \rho,~ \rho=E/V$.
First of all, we introduce  two  entropies
 for the holographic description of a universe~\cite{Verl}:
\begin{equation}
\label{2eq3}
  S_{\rm
 BV}=\frac{2\pi}{n}Ea,
  ~~ S_{\rm
 BH}=(n-1)\frac{V}{4G_{n+1}a},
\end{equation}
where $S_{\rm
 BV}$ and $S_{\rm
 BH}$ are called the Bekenstein-Verlinde and Bekenstein-Hawking entropies, respectively.
 We identify $S_{\rm BH}$ with the Bekenstein-Hawking entropy of a
 universe-size black hole, which is close to the
 usual expression of $A/4G_{n+1}$.
Then the first Friedmann equation can be rewritten as
\begin{equation}
\label{Fried} (Ha)^2=2\frac{S_{\rm BV}}{S_{\rm BH}}-k.
\end{equation}
 We  define a quantity $E_{\rm BH}$ which
corresponds to energy needed to form a universe-sized black hole:
$ S_{\rm BH}=(n-1)V/4G_{n+1}a \equiv 2\pi E_{\rm BH} a/n $. For
$Ha \le \sqrt{2-k}$, one finds the Bekenstein-Hawking bound for a
weakly self-gravitating system

\begin{equation}
\label{WB} E\le E_{\rm BH} \leftrightarrow S_{\rm BV} \le S_{\rm
BH},
\end{equation}
while  for $Ha \ge \sqrt{2-k}$,  a holographic bound is given for
a strongly self-gravitating system

\begin{equation}
\label{SB} E\ge E_{\rm BH} \leftrightarrow S_{\rm BV} \ge S_{\rm
BH}.
\end{equation}
The Bekenstein-Hawking bound is just a criterion on distinguishing
between  weakly  and strongly gravitating system. Making use of
the first Friedmann equation with $Ha=\sqrt{2-k}$, one finds that
$S_{\rm BV}= S_{\rm BH} \leftrightarrow E=E_{\rm BH}$. For $a\le
\sqrt{2-k}/H$, we have to use the energy bound $E \le E_{\rm BH}$
because its entropy bound $S_{\rm BV} \le S_{\rm BH}$ is less
important than the energy bound. This is  because $S_{\rm BV}$ is
not really  an entropy but rather  the energy. Also the role of
$S_{\rm BH}$ is not to serve a bound on the total entropy, but
rather on a sub-extensive component of the entropy. In this case,
we may introduce the original Bekenstein bound on the entropy
\begin{equation}
\label{OBB}S\le S_{\rm BV}.
\end{equation}
This bound is restricted within narrow limits. Also there is no
bound on the temperature of the system.  Hence we use mainly the
Bekenstein-Hawking energy bound for a weakly gravitating system
with  $Ha\le\sqrt{2-k}$.

For $Ha\ge\sqrt{2-k}$, one has $S_{\rm BV} \ge S_{\rm BH}$ and the
original Bekenstein  bound in Eq.(\ref{OBB}) should  be  replaced
by an appropriate one. A good candidate is the Hubble entropy
bound $S \le S_{\rm H}$. Here $S_{\rm H}$ is proportional to the
Bekenstein-Hawking entropy of a Hubble-size black hole ($H V_{\rm
H}/4G_{n+1}$) times the number of Hubble regions in the universe
($N_{\rm H}=V/V_{\rm H} \ge 1$)~\cite{Verl}. That is, one finds
that $S_{\rm H} \propto HV/4G_{n+1}$.  According to the
Fischler-Susskind-Bousso (FSB) proposal, the entropy flow $S$
through a contracting sheet is less to equal to $A/4G_{n+1}~(S \le
A/G_{n+1})$, where $A$ is the area of the surface from which the
light sheet originates. The infinitesimal form of this
prescription leads to the Hubble entropy bound  with the prefactor
$(n-1)$.

In order to derive the Hubble bounds explicitly, the Friedmann
equations (\ref{2eq2}) can be  cast to the cosmological
Cardy-Verlinde formula and cosmological Smarr formula,
respectively
\begin{eqnarray}
\label{2eq4}
 && S_{\rm H}=\frac{2\pi a}{n}\sqrt{E_{\rm
BH}(2E-kE_{\rm
BH})}, \nonumber \\
&& kE_{\rm BH}=n(E+pV -T_{\rm H} S_{\rm H}),
\end{eqnarray}
where the Hubble entropy ($S_{\rm H}$) and temperature ($T_{\rm
H}$) are defined by
\begin{equation}
\label{2eq5}
 S_{\rm H}=(n-1) \frac{HV}{4G_{n+1}},~~T_{\rm H}=-\frac{\dot H}{ 2\pi
 H}.
 \end{equation}
These are useful quantities for describing  a strongly gravitating
phase with  $Ha \ge \sqrt{2-k}$ and  are based on the holographic
principle and Friedmann equations. Actually, Eq.(\ref{2eq4})
corresponds to another representations of the two Friedmann
equations, which are expressed in terms of holographic quantities.

For a further study, we introduce a CFT-like radiation  whose
entropy and Casimir energy can be described by the Cardy-Verlinde
formula and the Smarr formula, respectively
\begin{eqnarray}
\label{2eq6} && S =\frac{2\pi L}{n}\sqrt{E_c(2E-E_c)},
  \nonumber \\
&& E_c=n(E+pV -T S),
\end{eqnarray}
where $S$ is the entropy of  CFT-like radiation living on an
$n$-dimensional space with size $L$ and  $E$ is the total energy
of CFT-like radiation. The first denotes  the entropy-energy
relation, while the second represents the relation between a
non-extensive part of the total energy (Casimir energy) and
thermodynamic quantities.  Here $E_c$ and $T$ stand for the
Casimir energy  and the temperature of radiation with
$\omega=1/n$. The above equations correspond to thermodynamic
relations for the CFT-radiation which are independent of the
Friedmann equations. Suppose that the entropy of  CFT-radiation in
the FRW universe can be described by the Cardy-Verlinde formula.
For general  $k$, a bound on the Casimir energy ($E_{\rm c} \le
E_{\rm BH}$) is equivalent to the Hubble entropy bound $S \le
S_{\rm H}$. It follows from Eqs.(\ref{2eq4}) and (\ref{2eq6}) that
the Hubble bounds for entropy and temperature are given
by\cite{Verl}
 \begin{equation}
 \label{2eq7}
 S \le S_{\rm H},~~ T \ge T_{\rm H}, ~~{\rm for}~ Ha \ge
 \sqrt{2-k}
 \end{equation}
which shows  two inequalities between geometric quantities
($S_{\rm H},T_{\rm H}$) and matter quantities ($S,T$). The Hubble
entropy bound can be saturated by the entropy of  CFT-radiation
filling the universe when its Casimir energy $E_c$ is large enough
to form a universe-size black hole.
 That is, if $kE_{\rm BH}=E_c$, one has  $S_{\rm
H}=S$ and $T_{\rm H}=T$. If it happens, equations (\ref{2eq4}) and
(\ref{2eq6}) coincide.
  This implies that the first Friedmann equation
somehow knows the entropy formula  for CFT-radiation filling the
universe.

Hereafter we use the Bekenstein-Hawking energy bound $E\le E_{\rm
BH}$ for a weakly gravitating system with $a\le \sqrt{2-k}/H$,
whereas for a strongly gravitating system with $a\ge
\sqrt{2-k}/H$, we use the Hubble bounds of $S \le S_{\rm H},~T\ge
T_{\rm H}$. Further the space dimensions is fixed to be  $n=3$.

\section{Holographic bounds on the inflationary universe}

A holographic bound with the UV cutoff can be applied to
describing the quantum fluctuations of inflation on sub-horizon
scales~\cite{HI}. The holographic principle provides a bound on
the UV cutoff scale of the effective theory of inflation. Let us
introduce the UV cutoff scale $\Lambda$  to parameterize our
ignorance of physics beyond this scale. This means that the cutoff
should be in physical momentum ($\wp <\Lambda$). In this case the
number of degrees of freedom depends on the UV cutoff. One usually
assumes that $\Lambda \propto M_p$. On sub-horizon scales the
energy density for scalar perturbations is expressed in terms of
the UV cutoff $\Lambda$ and the Hubble horizon $d_{\rm H}=1/H$
as~\cite{KS}
\begin{equation}
\label{den} \rho_{\rm qfi}=\frac{1}{32\pi^2} (\Lambda/d_{\rm
H})^2=\frac{1}{32\pi^2} (\Lambda H)^2,
\end{equation}
which shows that for an effectively flat spacetime ($k=0$), its
equation of state is given by a radiation $p=\rho/3$. In Section
5, we will obtain the  form  of $\rho_{\rm qfi}$ from the
Bekenstein-Hawking temperature bound. On sub-horizon scales, its
coarse-grained entropy is given by~\cite{GG}

\begin{equation}
\label{cg-en} S_{\rm  cg} \simeq \frac{(\Lambda d_{\rm
H})^2}{3\pi} =\frac{1}{3\pi}\Big(\frac{\Lambda}{H}\Big)^2
\end{equation}
which is the entropy for  relativistic particles at early time
before horizon crossing. Here there exist two candidates for the
entropy bound. One is the direct entropy bound from  the D-bound
that appears when a relativistic matter is embedded in  de Sitter
space~\cite{GH,Myung}. If one takes the  Hubble horizon of $1/H$
as the cosmological event horizon, then the increase of area is
given by
\begin{equation}
\label{area} \Delta A=A_{\rm f}-A_0= \frac{\epsilon_{\rm H}}{2}
A_{\rm f}.
\end{equation}
Here a Hubble slow-roll parameter $\epsilon_{\rm
H}=-\dot{H}/H^2<1$ is in the region $0<\epsilon_{\rm H}<1$ for
inflation, the initial horizon area $A_0$ for  de Sitter space
with the matter $\rho_{\rm qfi}$, and the final horizon area
$A_{\rm f}=4 \pi/H^2$ for de Sitter space after the matter is
excited into the cosmological horizon. Then the D-bound implies
that the entropy for the matter is limited by
\begin{equation}
\label{ent-b} S \le \frac{1}{4}\frac{\Delta
A}{G}=\frac{\pi}{2}\frac{\epsilon_{\rm
H}M_p^2}{H^2}=\frac{8\pi^2}{3}\frac{\rho_{\rm qfi}}{H^4}.
\end{equation}
Using the energy density in Eq.(\ref{den}), this leads to the
D-bound for quantum fluctuations of inflation~\cite{KS}

\begin{equation}
\label{ent-bb} S \le \frac{1}{12}\frac{\Lambda^2}{H^2}.
\end{equation}
Apparently the coarse-grained entropy in Eq.(\ref{cg-en}) violates
the  D-bound. Thus we cannot justify that the D-bound is satisfied
with  $S=S_{\rm cg}$ on sub-horizon scales. A relevant bound on
quantum fluctuations of inflation seems to be the energy bound.

If one uses the potential slow-roll parameter $\epsilon_{\rm
V}=1/2(V'/V)^2$, one has the indirect entropy bound. In this case
the change of entropy during one $e$-folding is limited by
\begin{equation}  \label{eebond}\Delta S \le
\frac{\pi}{2}\frac{\epsilon_{\rm V}M_p^2}{H^2}.
\end{equation}
On the other hand, the amount of entropy that exists the horizon
during one $e$-folding is given by $\Delta S = \Lambda^2/\pi H^2$.
Then  a bound on the UV cutoff  takes the form
\begin{equation}
\label{bcut}\Lambda \le \sqrt{8\pi^3\epsilon_{\rm V}}M_p.
\end{equation}
For an $e$-folding number $N=54(\sim 1/\epsilon_{\rm V})$, one
finds  $\Lambda \le 2M_p$. This means that the UV cutoff scale is
nearly of the same order of  the Planck scale $M_p$.

At late times after horizon crossing, the  entropy is described by
the entanglement entropy~\cite{KS}
\begin{equation}
\label{ent-en} S_{\rm  ent}=0.3 (\Lambda d_{\rm H})^2=0.3
\Big(\frac{\Lambda}{H}\Big)^2,
\end{equation}
which corresponds to the entropy for  classical fluctuations of
inflation outside the Hubble horizon (on super-horizon scales).
This is nearly of the same order of  the coarse-grained entropy
and scales like the area of $d_{\rm H}^2=1/H^2$. Now let us
introduce the Hubble entropy bound in Eq.(\ref{2eq7}) because the
system is in a strongly gravitating phase with $a\ge\sqrt{2}/H
\sim 1/H$. It is given by~\cite{Verl,Hogan}
\begin{equation}
S\le S_{\rm H}\sim N_{\rm H} \frac{M_p^2}{H^2}.
\end{equation}
Taking $\Lambda \sim M_p$, the Hubble entropy bound can be
rewritten as
\begin{equation}
S\le  N_{\rm H} \frac{\Lambda^2}{H^2}.
\end{equation}
It is evident from $a\ge 1/H$ that $N_{\rm H} \ge 1$. Then one
easily checks that the Hubble entropy bound is satisfied with the
entanglement entropy $S_{\rm ent}$.

We remark the Hubble temperature bound in Eq.(\ref{2eq7}).
Expressing the Hubble temperature $T_{\rm H}$ in terms of
$\epsilon_{\rm H}$ and the Gibbons-Hawking temperature $T_{\rm
GH}=\frac{H}{2 \pi}$~\cite{GH}, one finds the relation
\begin{equation}
\label{3eq12} T_{\rm H}=\epsilon_{\rm H}T_{\rm GH}.
\end{equation}
A condition for an accelerating universe ($0<\epsilon_{\rm H}<1$)
leads to  an  inequality
\begin{equation} T_{\rm H}
< T_{\rm GH}. \end{equation} This   is another representation for
inflation to occur, which is expressed in terms of the Hubble and
Gibbons-Hawking temperatures. Now we introduce the matter
temperature $T$ for classical fluctuations of inflation. The
holographic temperature bound implies
\begin{equation}
\label{3eq13} T \ge T_{\rm H}.
\end{equation}
We will show that this bound is valid for a slow-roll period of
inflation. We recall that this bound is derived from the
connection between the second Friedmann equation in
Eq.(\ref{2eq2}) and CFT-like radiation. For a slow-roll period of
inflation, the Hubble temperature is positive but less than the
Gibbons-Hawking temperature $T_{\rm GH}$.  For a de Sitter
inflation, one has $T_{\rm H}=0$ and  $ T_{\rm GH}>0$. This means
that at least the Hubble temperature is a limiting temperature to
define the matter temperature $T$ as $T\ge 0$ from the Hubble
temperature bound. For a de Sitter inflation, the holographic
temperature bound leads to the  definition of matter temperature:
$T\ge 0$ exactly. In the case of $T_{\rm GH}=T_{\rm H}$, inflation
comes to an end ($\epsilon_{\rm H}=1$). In the slow-roll inflation
there may be no cosmological horizon but one would still like to
describe the approximate thermal state.  Here we propose a
stronger constraint on the matter temperature than the Hubble
temperature bound
\begin{equation}
\label{3eqstro} T \ge T_{\rm GH}
\end{equation}
for classical fluctuations of inflation because the Hubble horizon
 plays a role of thermal heat bath with slowly varying
Gibbons-Hawking temperature $T_{\rm GH}=H/2\pi$.

We find that the  D-bound derived by analogy with de Sitter
spacetime is not appropriate for describing quantum fluctuations
of inflation with the UV cutoff $\Lambda$. This means that the
entropy bound is not suitable for a weakly gravitating system.
Instead, one expects that the energy bound will play an important
role. On the other hand, the indirect entropy bound leads to a
constraint on the UV cutoff scale which implies that $\Lambda \sim
M_p$.
 Fortunately, we show that the Hubble entropy and temperature
bounds are suitable for describing classical fluctuations of
inflation on super-horizon scales.

\section{Holographic energy bound}
Now we are in a position to ask  how  the cosmic holographic
bounds use to describe the present accelerating universe.
    For an effective
quantum field theory in a box of size  $L_{\rm \Lambda}$ with the
UV cutoff $\Lambda$, its entropy scales extensively as
 \begin{equation}
 \label{2eq8}
 S_{\rm \Lambda} \sim L_{\rm \Lambda}^3\Lambda^3,
 \end{equation}
 where  we choose  the volume
of the box as $V_{\rm \Lambda}=4\pi L_{\rm \Lambda}^3/3 \sim
L_{\rm \Lambda}^3$. This means that we start with the two
independent UV and IR cutoffs. Before we proceed, we comment on
the original Bekenstein bound of $S\le S_{\rm BV}$~\cite{Beke}.
Considering $S_{\rm BV}=2\pi E_{\rm \Lambda}L_{\rm \Lambda}\sim
(L_{\rm \Lambda}\Lambda)^4$,  this bound is  automatically
satisfied for an effective quantum field theory. Thus this bound
does not provides any constraint for an effective theory.

Considering the thermodynamics of black holes, one postulated that
the maximum entropy in the box of volume $V_{\rm \Lambda}$ behaves
non-extensively, growing only as the enclosed area $A_{\rm
\Lambda}$ of the box. If one limits the entropy to a holographic
bound
\begin{equation}
\label{BB} S_{\rm \Lambda} \sim L_{\rm \Lambda}^3\Lambda^3 \le
S_{\rm BH} \equiv\frac{2}{3}\pi M^2_p L_{\rm \Lambda}^2  \sim (M_p
L_{\rm \Lambda})^2,
\end{equation}
the effective theory can describe the black hole.  We call the
above the Bekenstein bound to compare with the original Bekenstein
bound of $S\le S_{\rm BV}$. Here the IR cutoff cannot be chosen
independently of the UV cutoff $\Lambda$.  It scales like $L_{\rm
\Lambda} \sim \Lambda^{-3}$. On the other hand, Cohen {\it et al.}
proposed another energy bound~\cite{CKN}
\begin{equation}
\label{EB} E_{\rm \Lambda} \sim L_{\rm \Lambda}^3\Lambda^4 \le
M_{\rm S} \sim L_{\rm \Lambda} M_p^2,
\end{equation}
where the IR cutoff scales as $L_{\rm \Lambda} \sim \Lambda^{-2}$.
This bound is more restrictive  than the Bekenstein bound in
Eq.(\ref{BB}).  The two scales $\Lambda$ and $L_{\rm \Lambda}$ are
independent to each other in the beginning. To reconcile the
breakdown of the quantum field theory to describe a black hole,
one proposes a relationship between UV and IR cutoffs. Then  we
have an effective field theory  with the IR cutoff which could
describe a system including even black holes.

Now we explain the bound of Eq.(\ref{EB}) within our framework.
Assuming that the present universe is in a weakly gravitating
phase, we can reinterpret it in view of the Bekenstein-Hawking
energy bound. From Eq.(\ref{WB}),  one finds that the holographic
bound is
 $E \sim L_{\rm
\Lambda}^3\Lambda^4 \le E_{\rm BH}\equiv 2L_{\rm \Lambda}M_p^2
\sim L_{\rm \Lambda}M^2_p$ with $ a \sim L_{\rm \Lambda}$. This
leads to the energy bound in Eq.(\ref{EB}). Also this inequality
can be derived from the Bekenstein-Hawking entropy bound of
$S_{\rm BV} \le S_{\rm BH}$. If $Ha=\sqrt{2}$, one finds the
saturation which means  that $S_{\rm BV}=S_{\rm BH}
\leftrightarrow E=E_{\rm BH}$. We remind the reader that $E_{\rm
BH}$ is an energy to form a universe-size black hole. The universe
is in a weakly self-gravitating phase when its total energy $E$ is
less than $E_{\rm BH}$, and in a strongly gravitating phase for
$E>E_{\rm BH}$. Comparing with the Bekenstein bound in
Eq.(\ref{BB}), the Bekenstein-Hawking bound comes out  when taking
both the Friedmann equation and holographic principle into
account.

 Consequently, the energy bound of Eq.(\ref{EB}) is
nothing but the Bekenstein-Hawking energy bound for a weakly
gravitating system. As is emphasized again, the energy is a
 rather important quantity for describing a weakly gravitating system than
 the entropy.

\section{Holographic dark energy}

In order to express the dark energy in terms of  the holographic
energy density, we take the largest $L_{\rm \Lambda}$ as the one
saturating the inequality of Eq.(\ref{EB}). Then  we obtain  a
holographic energy density,
\begin{equation} \label{hde}
\rho_{\rm \Lambda}=\frac{3c^2}{8\pi} \Big(\frac{M_p}{ L_{\rm
\Lambda}}\Big)^2
\end{equation}
with a numerical constant $3c^2$.
 In other words, one uses the equality $E=E_{\rm BH}$ of the Bekenstein-Hawking bound
 to get the holographic energy
density.  The Planck scale comes from this energy bound naturally.
The UV cutoff $\Lambda$ resolves into the IR cutoff $L_{\rm
\Lambda}$ and instead, the Planck scale $M_p$ plays a role of  the
UV cutoff.

We discuss the connection between Eq.(\ref{den}) and
Eq.(\ref{hde}).  For a while, we neglect the prefactors. These are
the same holographic form  except replacing $\Lambda$ and $d_{\rm
H}$ by $M_p$ and $L_{\rm \Lambda}$.   In the case of quantum
fluctuations $\rho_{\rm qfi}$ of inflation  on sub-horizon scales,
the UV cutoff $\Lambda$ is determined to be the Planck scale $M_p$
by the indirect entropy bound.  On the other hand, one can choose
the IR cutoff $L_{\rm \Lambda}$ to be the Hubble horizon because
of $a\le \sqrt{2}/H\sim 1/H$ on sub-horizon scales. Then, we
obtain the energy density  $\rho_{\rm qfi}$ for quantum
fluctuations of inflation from the holographic energy density
$\rho_{\rm \Lambda}$. Furthermore, we note that  the
coarse-grained, entanglement, and Hubble entropies take the same
holographic form: $S \sim (M_pd_{\rm H})^2$.  From the Bekenstein
entropy bound in Eq.(\ref{BB}), one finds  the same form of
holographic   entropy: $S_{\rm \Lambda}\sim (M_pL_{\Lambda})^2$. A
difference is between $d_{\rm H}=1/H$ in the past and
$L_{\Lambda}$ in the present. These show the close relationship
between two weakly gravitating systems.

 Here one has three possibilities for
$L_{\rm \Lambda}$ to describe the dark energy. If one chooses the
IR cutoff as the size of our universe ($L_{\rm \Lambda}=d_{\rm
H}$), the resulting energy density is comparable to the present
dark energy\cite{HMT}. Even though this holographic approach leads
to the data, this description is incomplete because it fails to
explain the present universe  with $ \omega \le -0.78$~\cite{HSU}.
Explicitly, the Friedmann equation including  a matter  of
$\rho_m$ is then given by $\rho_m=3(1-c^2)M_p^2H^2/8\pi$, which
shows that the equation of state for the dark energy is given by
$\omega_{\rm d}=0$. However, an accelerating universe requires
$\omega <-1/3$ and thus it is not the case.
 To
resolve this situation, one is forced to introduce the particle
horizon $L_{\rm \Lambda}=R_{\rm H}=a \int_0^t (dt/a)$ which was
used in the holographic description of cosmology by Fischler and
Susskind~\cite{FS}.  The Friedmann equation of $H^2=8\pi \rho_{\rm
\Lambda}/3M_p^2$ leads to an integral equation $HR_{H}=c$, which
gives $\rho_{\rm \Lambda}\sim a^{-2(1+1/c)}$. Unfortunately it
implies a decelerating universe with $\omega_{\rm H}=
-1/3(1-2/c)>-1/3$. In order to find an accelerating universe, we
consider  quantum fluctuations of inflation on sub-horizon scales.
In this case, the Hubble horizon plays a role of the event horizon
approximately~\cite{BOU3}. Similarly we choose the future event
horizon $R_{\rm h}=a \int_t^{\infty} (dt/a)$ as the IR cutoff
$L_{\rm \Lambda}$~\cite{LI}. Using the Friedmann equation of
$HR_{\rm h}=c$, one finds $\rho_{\rm \Lambda}\sim a^{-2(1-1/c)}$
with $\omega_{\rm h}= -1/3(1+2/c)<-1/3$.  For $c=1$, we recover a
case of cosmological constant with $\omega_{\rm h}=-1$.

\section{Summary}

First of all we find that the  D-bound derived by analogy with de
Sitter spacetime is not appropriate for describing quantum
fluctuations of inflation with the UV cutoff $\Lambda$. This means
that the entropy bound is not suitable for a weakly gravitating
system. Instead, one expects that the energy bound will play an
important role. On the other hand, the indirect entropy bound
leads to a constraint on the UV cutoff scale which implies that
$\Lambda \sim M_p$. Fortunately, we show that the Hubble entropy
and temperature bounds are suitable for describing classical
fluctuations of inflation on super-horizon scales with $\Lambda
\sim M_p$. This is because the classical fluctuations of inflation
corresponds to a strongly gravitating system with $a\ge
\sqrt{2}/H$.

Furthermore, the energy bound of Eq.(\ref{EB}) is nothing but the
Bekenstein-Hawking energy bound for a weakly gravitating system.
In this case the energy is a
 rather suitable quantity for describing a weakly gravitating system than
 the entropy. The holographic energy density $\rho_{\rm
 \Lambda}$
 is obtained  from the one saturating the Bekenstein-Hawking energy
 bound. The energy density $\rho_{\rm qfi}$ for quantum fluctuations of inflation
 is closely related to the holographic energy density $\rho_{\rm
 \Lambda}$ for the dark energy.

In order to describe the dark energy in terms of the holographic
energy density, one has to choose the IR cutoff $L_{\rm \Lambda}$.
In this approach  we don't know its equation of state exactly
before solving the first Friedmann equation of  $HL_{\rm
\Lambda}=c$. For quantum fluctuations of inflation on sub-horizon
scales, we know that they behave a radiation with $\omega=1/3$.
However, in the holographic approach to a weakly gravitating
system, there exists information loss. This arises because one
does not use the second Friedmann equation. As is shown in
Eqs.(\ref{2eq2}) and (\ref{2eq4}), this contains information on
the equation of state for a matter partly. In the case of $k=0$,
one obtains $(\rho+p)V=T_{\rm H}S_{\rm H}$ which shows that
equation of state depends on the Hubble temperature $T_{\rm H}$.
In the case of $T_{\rm H}= 0$, one obtains $\omega = -1$ which is
the de Sitter universe with a cosmological constant. In the case
of $T_{\rm H} \not= 0$, one finds  $p=-\rho+ (S_{\rm H}/V)T_{\rm
H}>-\rho$ which implies that $\omega>-1$. This means that the
geometric quantities of $S_{\rm H},~T_{\rm H}$ and $V$ may
determine the  equation of state.

\begin{table}
 \caption{Summary for three systems: quantum fluctuations of inflation (qfi) on sub-horizon scales,
 classical fluctuations  of inflation (cfi) on super-horizon scales, and dark energy (de) in the future.
 The two cutoff scales are introduced: the UV cutoff scale is the
 Planck scale ($\Lambda \sim M_p)$ for all three cases, but the IR cutoff
 scale depends on the system. Explicitly, $L_{\rm \Lambda}=1/H$, for qfi and cfi and $L_{\rm \Lambda}=R_{\rm h}$ for
 de.}
\begin{tabular}{|c|c|c|c|}\hline
     & qfi  & cfi & de \\ \hline
  type of system & weakly  & strongly  & weakly  \\\hline
  type of bound & $E\le E_{\rm BH}$ & $S\le S_{\rm H},~T\ge T_{\rm H}$ & $E\le E_{\rm BH}$ \\\hline
  energy density &$\rho_{\rm qfi}=\frac{1}{32\pi^2}(\Lambda H)^2$&
  $\rho_{\rm cfi} \sim (\Lambda H)^2$
  & $\rho_{\rm \Lambda}=\frac{3c^2}{8\pi}\Big(\frac{M_p}{R_{\rm h}}\Big)^2$\\\hline
  entropy& $S_{cg}=\frac{1}{3\pi}\Big(\frac{\Lambda}{H}\Big)^2$ & $S_{en}=0.3\Big(\frac{\Lambda}{H}\Big)^2$
  & $S_{\rm \Lambda}=(M_p R_{\rm h})^3$ \\\hline
  equation of state & $p=\rho/3$ & $ p \sim \rho/3$ & $ p =-\rho$ for $c=1$\\ \hline
 \end{tabular}
 \end{table}

As is shown in Table 1, the two combinations of the UV and IR
cutoffs ($\Lambda/L_{\rm \Lambda},~\Lambda L_{\rm \Lambda}$)
determine all holographic energy densities and  holographic
entropies. In the case of $\rho_{\rm cfi}$, even though its
correct form is not known, one conjectures  its form $\rho_{\rm
cfi} \sim (\Lambda H)^2$. As a result, there exist a close
connection between three systems. Other approaches to the
connection between inflation and dark energy  appeared in
ref.~\cite{Pad1}.

In conclusion the cosmic holographic bounds with the UV and IR
cutoffs could describe weakly gravitating systems with $a\le
\sqrt{2}/H$. The energy bound is very useful for quantum
fluctuations of inflation in the past and the dark energy in the
future. In these cases, the UV cutoff is the Planck scale and the
IR cutoff  is the Hubble horizon for quantum fluctuations of
inflation and the future event horizon for the dark energy. For
classical fluctuations of inflation on super-horizon scales, we
use the Hubble bounds to obtain the lower limit of its entropy and
the upper limit of its temperature. Here the UV cutoff  is also
the Planck scale and the IR scale is the Hubble horizon on
super-horizon scales.

\section*{Acknowledgment}
This work  was supported in part by KOSEF, Project Number: R02-2002-000-00028-0.


\begin{thebibliography}{99}
\bibitem{SN} S. J. Perllmutter {\it et al.}, Astrophys. J. {\bf 517},
565(1999)[astro-ph/9812133
]; 
A. G. Reiss {\it et al.}, Astron. J. {\bf 116},
1009 (1998)[astro-ph/9805201 ]; 
A. G. Reiss  {\it et al.}, Astrophys. J. {\bf 607},
665(2004)[astro-ph/0402512]; 

\bibitem{Wmap} H. V. Peiris  {\it et al.}, Astrophys. J. Suppl. {\bf 148} (2003) 213
[astro-ph/0302225];
C. L. Bennett  {\it et al.}, Astrophys. J. Suppl. {\bf 148} (2003)
1[astro-ph/0302207];
D. N. Spergel  {\it et al.}, Astrophys. J. Suppl. {\bf 148} (2003)
175[astro-ph/0302209].

\bibitem{Inf} A.~H. ~Guth,
Phys.\ Rev.\ D {\bf 23}, 347 (1981);
A.~D.~Linde, Phys.\ Lett.\ B {\bf 108}, 389 (1982);
A.~Albrecht and P.~J.~Steinhardt, Phys.\ Rev.\ Lett.\  {\bf 48},
1220 (1982).

\bibitem{Beke}J.~D.~Bekenstein,
Phys.\ Rev.\ D {\bf 23}, 287 (1981).


\bibitem{KS} E. Keski-Vakkuri and M. S. Sloth, JCAP {\bf 0308}, 001
(2003) [hep-th/0306070].

\bibitem{CKN} A. Cohen, D. Kaplan, and A. Nelson, Phys. Rev. Lett.
{\bf 82}, 4971 (1999)[arXiv:hep-th/9803132].

\bibitem{HMT} P. Horava and D. Minic, Phys. Rev. Lett.
{\bf85}, 1610 (2000)[arXiv:hep-th/0001145];
S. Thomas, Phys. Rev. Lett. {\bf 89}, 081301 (2002).



\bibitem{HSU} S. D. Hsu, Phys. Lett. B {\bf 594},13
(2004)[hep-th/0403052].

\bibitem{WM} P. Wang and X. Meng, Class. Quant. Grav. {\bf 22}, 283(2005)
[astro-ph/0408495].

\bibitem{HOV} R. Horvat, Phys. Rev. D {\bf 70}, 087301 (2004)
[astro-ph/0404204].


\bibitem{LI} M. Li, Phys. Lett. B {\bf 603}, 1
(2004)[hep-th/0403127].

\bibitem{FEH}
Q-C. Huang and Y. Gong, JCAP {\bf 0408}, 006
(2004)[astro-ph/0403590];
Y. Gong, Phys. Rev. D {\bf 70}, 064029 (2004)[hep-th/0404030];
B. Wang, E. Abdalla and Ru-Keng Su, hep-th/0404057;
S. Hsu and  A. Zee, hep-th/0406142;
P. F. Gonzalez-Diaz, hep-th/0411070;
S. Nobbenhuis, gr-qc/0411093.


\bibitem{Myung2} Y. S. Myung, hep-th/0412224.


\bibitem{Hooft}G.~'t Hooft,
gr-qc/9310026;
L.~Susskind,
J.\ Math.\ Phys.\  {\bf 36}, 6377 (1995) [hep-th/9409089].

\bibitem{FS}W.~Fischler and L.~Susskind,
hep-th/9806039.

\bibitem{Hubb}
R.~Easther and D.~A.~Lowe,
Phys.\ Rev.\ Lett.\  {\bf 82}, 4967 (1999) [hep-th/9902088];
G.~Veneziano,
Phys.\ Lett.\ B {\bf 454}, 22 (1999) [hep-th/9902126];
G.~Veneziano,
hep-th/9907012;
R.~Brustein and G.~Veneziano,
Phys.\ Rev.\ Lett.\  {\bf 84}, 5695 (2000) [hep-th/9912055];
D.~Bak and S.~J.~Rey,
Class.\ Quant.\ Grav.\  {\bf 17}, L83 (2000) [hep-th/9902173];
N.~Kaloper and A.~D.~Linde,
Phys.\ Rev.\ D {\bf 60}, 103509 (1999) [hep-th/9904120].

\bibitem{Bous}R.~Bousso,
JHEP {\bf 9907}, 004 (1999) [hep-th/9905177];
R.~Bousso,
JHEP {\bf 9906}, 028 (1999) [hep-th/9906022].

\bibitem{Verl}E.~Verlinde,
hep-th/0008140.

\bibitem{SV}
B. Wang, E. Abdalla, and  T. Osada, Phys. Rev. Lett. {\bf 85},
5507 (2000) [astro-ph/0006395];
 S.  Nojiri and  S. D. Odintsov,
Int. J. Mod. Phys. A {\bf 16}, 3273 (2001)[hep-th/0011115];
I.~Savonije and E.~Verlinde,
Phys.\ Lett.\ B {\bf 507}, 305 (2001) [hep-th/0102042];
S.  Nojiri and  S. D. Odintsov, Class. Quant. Grav. {\bf 18}, 5227
(2001)[hep-th/0103078];
R. G.  Cai, Y. S.  Myung, and  N. Ohta, Class. Quant. Grav. {\bf
18}, 5429 (2001) [hep-th/0105070];
S.  Nojiri, S. D. Odintsov, and S. Ogushi,  Int. J. Mod. Phys. A
{\bf 17}, 4809 (2002)[hep-th/0205187];
R.~G.~Cai and Y.~S.~Myung, Phys.\ Rev.\ D {\bf 67}, 124021
(2003)[arXiv:hep-th/0210272];
Y.~S.~Myung, Phys.\ Lett.\ B {\bf 578}, 7 (2004)
[arVix:hep-th/0306180];
D.~Youm, Phys.\ Lett.\ B {\bf 515}, 170 (2001)
[arXiv:hep-th/0105093];
 M. R. Setare and  R. Mansouri,
Int. J. Mod. Phys. A {\bf 18}, 4443 (2003)[hep-th/0210252];

\bibitem{HI} N. Kaloper, M. Kleban, A. Lawrence, and S. Shenker,
Phys. Rev. D {\bf 66}, 123510 (2002) [hep-th/0201158];
U. H. Danielsson, Phys. Rev. D {\bf 66}, 023511 (2002)
[hep-th/0203198];
 V. Bozza, M. Giovannini, and G. Veneziano
JCAP {\bf 0305}, 001 (2003) [hep-th/0302184].

\bibitem{GG} M. Gasperini and  M. Giovannini,
 Phys. Lett. B {\bf 301}, 334 (1993) [gr-qc/9301010].


\bibitem{GH}G.~W.~Gibbons and S.~W.~Hawking,
Phys.\ Rev.\ D {\bf 15}, 2738 (1977).



\bibitem{Myung}  V. Balasubramanian, J. de Boer, and D. Minic, Phys. Rev.  {\bf
D65} (2002) 123508 [hep-th/0110108];
Y. S. Myung, Mod. Phys. Lett. {\bf A16} (2001) 2353
[hep-th/0110123];
R. G. Cai, Y. S. Myung, and Y. Z. Zhang, Phys. Rev. {\bf D65}
(2002) 084019 [hep-th/0110234];
 A. M. Ghezelbash and  R.B. Mann, JHEP {\bf 0201}, 005 (2002) [hep-th/0111217].

\bibitem{Hogan} C. J. Hogan, astro-ph/0406447.



\bibitem{BOU3} R. Bousso, hep-th/0412197.


\bibitem{Pad1} K. Enqvist and M. S. Sloth, Phys. Rev. Lett. {\bf 93}, 221302 (2004)
[hep-th/0406019];
J. D. Bjorken, astro-ph/0404233;
 T. Padmanabhan, astro-ph/0411044.



\end{thebibliography}
\end{document}